\documentclass[traditabstract]{aa}
\usepackage{txfonts}
\usepackage{graphicx}
\usepackage[]{natbib}
\usepackage{mathrsfs}
\bibpunct{(}{)}{;}{a}{}{,}

\usepackage{rotating}
\usepackage[colorlinks=true, breaklinks=true, linkcolor=blue, citecolor=blue, urlcolor=blue]{hyperref}

\newcommand{\coa}[0]{\mbox{CoRoT-2A}}
\newcommand{\cocapa}[0]{\mbox{CoRoT-2A}}
\newcommand{\cob}[0]{\mbox{CoRoT-2b}}
\newcommand{\co}[0]{\mbox{CoRoT-2}}
\newcommand{\chan}[0]{\textit{Chandra}}

\newcommand{\cahk}[0]{\ion{Ca}{ii} \mbox{H and K}}
\newcommand{\cahkabs}[0]{Ca~H$_4$~and~K$_4$}
\newcommand{\teff}[0]{$T_\mathrm{eff}$}
\newcommand{\tmass}[0]{\mbox{2MASS~J19270636+0122577}}
\newcommand{\ergcms}[0]{\mbox{erg\,cm$^{-2}$\,s$^{-1}$}}
\newcommand{\ergs}[0]{\mbox{erg\,s$^{-1}$}}
\newcommand{\kms}[0]{\mbox{km\,s$^{-1}$}}

\begin{document}

\title{The corona and companion of \coa. Insights from X-rays and optical spectroscopy\thanks{Based on observations obtained with UVES at the ESO VLT Kueyen telescope (program ID 385.D-0426) and the \chan \ X-ray Observatory (obs. ID 10989).}}
\author{S. Schr\"oter \and S. Czesla \and U. Wolter \and H.M. M\"uller \and K.F. Huber \and J.H.M.M. Schmitt}
\institute{Hamburger Sternwarte, Universit\"at Hamburg, Gojenbergsweg 112, 21029 Hamburg, Germany}
\date{}
\abstract
{
\co \ is one of the most unusual planetary systems known to date.
Its host star is exceptionally active,
showing a pronounced, regular pattern of optical variability caused by magnetic activity. 
The transiting hot Jupiter, \cob, shows one of the largest known radius anomalies.
We analyze the properties and activity of \coa \ in the optical
and X-ray regime by means of a high-quality UVES spectrum and a $15$~ks \chan \ exposure both
obtained during planetary transits.
The UVES data are analyzed using various complementary methods of high-resolution stellar spectroscopy.
We characterize the photosphere of the host star by deriving accurate stellar parameters such as effective temperature, surface gravity, and abundances. Signatures of stellar activity, Li~abundance, and interstellar absorption are investigated to provide constraints on the age and distance of \co. 
Furthermore, our UVES data confirm the presence of a late-type stellar companion to \coa \ that is gravitationally bound to the system.
The \chan \ data provide a clear detection of coronal X-ray emission from \coa, for which we obtain an
X-ray luminosity of $1.9\times10^{29}$~\ergs.
The potential stellar companion remains undetected in X-rays.
Our results indicate that the distance to the \co \ system is $\approx 270$~pc, and the most
likely age lies between $100$ and $300$~Ma. Our X-ray observations show that the planet is immersed in an intense field of high-energy radiation.
Surprisingly, \coa's likely coeval stellar companion, which we find to be of late-K spectral
type, remains X-ray dark. Yet, as a potential third body in the system, the companion could account for \cob's slightly eccentric orbit.
}
\maketitle

\section{Introduction}

The \co \ system stands out of the plethora of known exoplanet systems both for its exceptionally active host star and its
unusually inflated planet.
The hot Jupiter \cob \ is the second transiting planet discovered by the space-based CoRoT mission \citep{Alonso2008}; its planetary nature was confirmed by spectroscopic follow-up observations with SOPHIE and HARPS \citep{Bouchy2008}.
The planet orbits its host star every $1.74$~days. Given its mass of $3.31$~$M_\mathrm{J}$ and large radius of $1.465$~$R_\mathrm{J}$ \citep{Alonso2008}, \cob \ appears to be anomalously inflated in comparison to current evolutionary models \citep{GuillotHavel2011}.
A spectral analysis showed that its host star, \coa, is a G7 dwarf with solar composition. Its
spectrum shows strong Li~I absorption and emission-line cores in \cahk, indicating that \coa \ is a young and active star \citep{Bouchy2008}.
\coa \ has a close visual companion, \tmass, separated by about $4\arcsec$.
Photometric magnitudes from the optical to the infrared concordantly suggest that this object is a late-K or early-M type star located at the same distance as \co \ \citep{Alonso2008,Gillon2010}. Thus, \coa \ and its visual companion possibly form a physical pair.

The continuous photometric data of \coa \ provided by the CoRoT telescope span $152$~days.
\coa's light curve shows a distinct pattern of variability caused by starspots. In several studies, the light curve was
used to reconstruct the surface brightness distribution of \coa: \citet{Lanza2009} applied a light-curve inversion technique and found that most spots are concentrated in two active longitudes of alternating strength located on opposite hemispheres. Moreover, it was demonstrated that starspots influence the profiles of transit light-curves and that this effect cannot be neglected in transit modeling \citep{Wolter2009, Czesla2009}.
Because the latitudinal band eclipsed by the planet is accurately known \citep{Bouchy2008}, it is even feasible to study the spot coverage on the surface section recurrently eclipsed by the planet \citep{Huber2009, Silva-Valio2010, Huber2010}.

Secondary eclipses of \cob \ were observed in the optical with CoRoT \citep{Alonso2009, Snellen2010},
in the infrared with Spitzer \citep{Gillon2010, Deming2011},
and from the ground \citep{Alonso2010}.
While atmospheric models \citep{Fortney2008} suggest the presence of a
stratospheric thermal-inversion layer in \cob \ caused by the
strong irradiation, the observational situation remains inconclusive \citep[e.g.,][]{Deming2011}.
The observed emission of the planet is currently incompatible with any kind of standard atmosphere model, and
more sophisticated approaches including, for example, substantial carbon monoxide mass loss or
additional substructure in the atmosphere
may be needed to explain the observations \citep[see][for a discussion]{Deming2011}; the substantial activity
of the host star adds another complicating factor to the picture \citep[e.g.,][]{Knutson2010}.

Using KECK/HIRES data, \citet{Knutson2010} searched for a relation between stellar activity as manifested by chromospheric emission in the \cahk \ line cores and the emission spectra of hot Jovians. Among their sample of planet host-stars, \coa \ stands out as being the most active as measured by its $\log{R'_{\mathrm{HK}}}$ index.
\citet{AmmlervonEiff2009} reanalyzed the archival UVES data presented in \citet{Bouchy2008} (program 080.C-0661D) and determined precise estimates of \coa's spectral properties such as effective temperature and iron abundance.
In August 2009, \citet{Gillon2010} obtained and analyzed a new UVES spectrum (program 083.C-0174C). The authors provide a more detailed discussion of the Li absorption line and derive an age between 30 and 316~Ma for \coa.
Additionally, they find evidence for a slight eccentricity of $0.014 \pm 0.008$ for the planetary orbit,
which they attribute to the youth of the system, if not caused by \coa's potential stellar companion.

\citet{GuillotHavel2011} investigated the matter of \coa's anomalously inflated planet on theoretical grounds by simultaneously modeling the planetary and stellar evolution including stellar activity. The authors' models favor two classes of solutions: Either a young system with a star on the pre-main sequence ($30-40$~Ma) or a much older system ($>100$~Ma) with a main-sequence star. The authors discuss several effects that could have led to the anomalously large radius of \cob. While they argue that the influence of starspots is minor, either the presence of additional infrared opacity sources in the planetary atmosphere that reduces the rate of heat loss during the planet's evolution or a recent interaction with a third body in the system that leaves the planet in an eccentric orbit could account for the observed radius anomaly.

In this work, we present a new UVES spectrum of \coa \ and its visual companion, which we analyzed in detail to refine the spectroscopic parameters and chromospheric activity indicators. We especially aim at deriving several independent estimates of the age and the distance of the \co \ system.
First, we present the results of our analysis of the high-resolution UVES spectrum of \coa \ and our low-quality spectrum of its visual companion.
Second, we present our analysis of the first X-ray observation of \coa.
We proceed by discussing the physical implications of our findings (Sect.~\ref{sec:Dicussion}) and,
finally, we present our conclusions.

\section{Optical spectroscopy with UVES}

\subsection{Observations}
\label{sec:Observations}
On June 7, 2010, we acquired 24 high-resolution spectra of \coa \ with the UVES spectrograph \citep{Dekker2000} mounted at the VLT Kueyen telescope (program ID 385.D-0426(A)). 
The instrument was set up in
``Dichroic 2'' mode with a slit width of $0.7\arcsec$.
We used the cross-disperser \#4 with a central wavelength of $7600$~\AA \ providing a wavelength coverage of
$3800-5000$~\AA \ on the blue arm and $5700-9500$~\AA \ on the red arm.
Around $7600$~\AA \ a section of $100$~\AA \ is missing because of the gap
between the two detectors.
Because no overbinning was applied during CCD read-out, we
reach a spectral resolving power of about $60\,000$.

In this set-up, we obtained $24$ individual spectra with exposure times of $800$~s each. The observations were
scheduled to cover a full planetary transit of \cob \ including a reasonable time span before and after
the actual transit event.
Owing to worsening seeing conditions, the last 13 observations were carried out using the image slicer to
reduce light losses. Additionally, we used an archived UVES spectrum obtained on Oct. 13, 2007 in the framework of the program
080.C-0661(D) for comparison.

To reduce the UVES echelle data, we applied the UVES pipeline
in version 4.7.8 with its associated standard recipes
and the REDUCE package developed by \citet{PiskunovValenti2002}.
Our analysis is based on the REDUCE spectra unless stated otherwise.
Background-sky subtraction for exposures taken with the image slicer is difficult
because there is hardly any sky area left on the detector that could be used to extract the background. Indeed,
the UVES pipeline does currently not apply any such subtraction for exposures taken with the image slicer.
The $24$ individual UVES spectra were combined and yielded a high-resolution spectrum of \coa \ with a signal-to-noise ratio (SNR) of about $200$ at $6500$~\AA.
Although sky emission-lines are present in the spectrum, they did not affect our analysis.

\subsection{Analysis of optical data}
\label{sec:Analysisofopticaldata}

The stellar spectrum conveys information about the physical conditions
in the stellar atmosphere. The parameters of primary interest in our analysis were the effective temperature,
elemental abundances, surface gravity, and microturbulence velocity.
Unfortunately, those parameters cannot be determined independently from the observed stellar spectrum, but they are
highly correlated.
Several techniques are commonly used in spectroscopic analyses, and
the results depend on the underlying assumptions and implementation.
Systematic errors can originate in the data reduction process or in differences in the adopted
atomic data.
As a consequence, error bars based on purely statistical considerations usually underestimate
the true uncertainty. Therefore, we applied a number of independent analyses to corroborate
the validity of our parameter and error estimates.
In the following analysis, we concentrate on the time-averaged UVES spectrum of \coa. 
An analysis of the
temporally resolved properties will be presented in another context.

\subsection{Elemental abundances via excitation/ionization balance}
\label{sec:ExcitationIonization}

\begin{table*}[t!]
  \begin{minipage}[h]{\textwidth}
    \renewcommand{\footnoterule}{}
    \caption{Stellar parameters and iron abundance of \coa \ determined from the analysis of Fe~I~and~II lines, the \cite{Sousa2007} line-ratio technique, and SME line profile fitting. \label{tab:CoRoT2specPars}}
    \begin{center}
      \begin{tabular}{l c c c c c l}
      \hline \hline
      Stellar parameters & \teff \ [K] & $\log{g}$ (cgs) & $\xi_t$ [km\,s$^{-1}$] & [Fe/H] & $N$(Fe~I, Fe~II) & Source \\
      \hline
      Excitation/Ionization balance & $5598\pm34$ & $4.47\pm0.14$ & $1.75\pm0.04$ & $0.04\pm0.02$ & $212,26$ & this work (Sect.~\ref{sec:ExcitationIonization}) \\
      Excitation/Ionization balance & $5608\pm37$ & $4.71\pm0.20$ & $1.49\pm0.06$ & $0.07\pm0.04$ & $26$, $9$ & \cite{AmmlervonEiff2009} \\
      \ion{Fe}{i}~and~\ion{}{ii} lines             & $5625\pm120$ & $4.3\pm0.2$ & $0.0\pm0.1$ & & & \cite{Bouchy2008} \\
      Line ratio calibration & $5513\pm111$ & & & & & this work (Sect.~\ref{sec:LineRatios})\\
      SME global fit & $5475\pm44$ & $4.62\pm0.06$ & $1.52$ & $-0.06\pm0.03$  & & this work (Sect.~\ref{sec:SME})\\
      H$\alpha$ $\lambda 6563$ & $5510^{+90}_{-70}$ & & & & & this work (Sect.~\ref{sec:SMEteff})\\
      H$\alpha$ $\lambda 6563$ & $5450\pm120$ & & & & & \cite{Bouchy2008}\\
      \ion{Ca}{i} $\lambda 6122$, $\lambda 6162$, $\lambda 6439$ & & $4.49\pm0.14$ & & & & this work (Sect.~\ref{sec:SMElogg})\\
      Na~D $\lambda 5890$, $\lambda 5896$ & & $4.53\pm0.18$ & & & & this work (Sect.~\ref{sec:SMElogg})\\
      \hline
    \end{tabular}
    \end{center}
  \end{minipage}
\end{table*}

In our high-resolution UVES spectrum of \coa, we measured equivalent widths (EWs) of lines of neutral (\ion{Fe}{i}) and ionized iron (\ion{Fe}{ii}) and several other metals (Na, Mg, Al, Si, Ca, Ti, V, Cr, Mn, Co, Ni, and Ba). Our selection of lines is based on the line lists provided by \citet{Sousa2008} and \citet{BubarKing2010}, out of which we compiled a list of iron lines without severe blends with excitation potentials below 5~eV and EWs between $10$~and $200$~m\AA. Our resulting line list comprises 212 \ion{Fe}{i} lines, 26 \ion{Fe}{ii} lines, and 162 lines of other metals.

To measure the EWs of a large number of spectral lines, \cite{Sousa2007} developed the ARES\footnote{see \url{http://www.astro.up.pt/~sousasag/ares/}.} code. This algorithm detects spectral lines by evaluating numerical derivatives of the spectrum. For the following analysis, we set up our own implementation of the ARES algorithm, extending it at several points, for example, by a low-pass Fourier filter to suppress noise effects and an estimation of the local continuum. Our tool runs semi-automatically, allowing the user to interactively improve the fit result by visual inspection where desired. The normalization of the spectrum is made manually by comparing the observed to a synthetic spectrum to identify regions of undisturbed continuum, which are then used as nodes for a linear (or cubic) spline fit.

The EWs thus obtained were used as input for the 2010 version of MOOG\footnote{see \url{http://www.as.utexas.edu/~chris/moog.html}.} \citep{Sneden1973} together with ATLAS plane-parallel model atmospheres\footnote{see \url{http://kurucz.harvard.edu/grids.html}.} \citep{Kurucz1993}. We used MOOG to derive the effective temperature by minimizing the correlation between iron abundance and excitation potential, whereas the microturbulence velocity was obtained by removing the correlation with reduced equivalent width, i.e., the EW normalized by the central wavelength of the line. The surface gravity is derived by minimizing the difference between the resulting \ion{Fe}{i} and \ion{Fe}{ii} abundances.

To account for possible errors in the atomic line parameters, the spectroscopic analysis proceeds differentially to the Sun. Therefore, we measured each EW in both a lunar UVES spectrum provided by the ESO Quality Control and Data Processing Group\footnote{see \url{http://www.eso.org/observing/dfo/quality/}} and the stellar spectrum and subsequently subtracted the solar abundance from the resulting stellar abundance. We used the freely available PYSPEC\footnote{see \url{http://www.pas.rochester.edu/~ebubar/speclink.html}} Python interface to MOOG by E.~Bubar for our differential spectroscopic analysis \citep[for details, see][]{BubarKing2010}. The abundances of the other metals were derived in the same way, but keeping the input stellar atmosphere model fixed.

We present our results in Table~\ref{tab:CoRoT2specPars}. The listed errors on effective temperature and microturbulent velocity are estimated by investigating the
correlation of the iron abundance with excitation potential and reduced equivalent width, respectively, as measured by Spearman's rank correlation coefficient. In particular,
the parameters were varied until a $1\sigma$ correlation was found, and the associated values were then
used as an estimator of the reasonable
parameter range. The error on $\log{g}$ was propagated based on the errors of the remaining parameters as detailed in \cite{BubarKing2010}.
Our spectral parameters agree well with earlier results obtained by \cite{Bouchy2008} and \cite{AmmlervonEiff2009} from the analysis of \ion{Fe}{i}~and~\ion{}{ii} lines. 

The inferred elemental abundances are summarized in Table~\ref{tab:CoRoT2elemAbund}. Neither for Ni nor for \ion{Ti}{i}~and~\ion{}{ii} did we find any strong correlations of abundance with excitation potential or equivalent width given the atmospheric model derived from the \ion{Fe}{i}~and~\ion{}{ii} lines. This indicates that the stellar parameters were indeed correctly chosen. Within the errors the elemental abundances are compatible with the solar values. The overabundance of \ion{Ba}{ii} is a result of the lines being blended with Fe~lines. We therefore redetermined the \ion{Ba}{ii} abundance via line synthesis and obtained a value of $+0.13\pm0.09$~dex, which better agrees with the remaining elemental abundances.

\begin{table}[t!]
  \begin{minipage}[h]{0.5\textwidth}
    \renewcommand{\footnoterule}{}
    \caption{Elemental abundances for \coa \ relative to the Sun with the number of lines $N$(X)
             used for each element. \label{tab:CoRoT2elemAbund}}
    \begin{center}
      \begin{tabular}{l c r}
      \hline \hline
      Elem. & [X/H] & $N$(X) \\
      \hline
      \ion{Mg}{i} & $-0.16\pm0.13$ & 4\\
      \ion{Si}{i} & $-0.05\pm0.11$ & 17\\
      \ion{Ca}{i} & $+0.12\pm0.10$ & 14\\
      \ion{Ti}{i} & $+0.05\pm0.14$ & 25\\
      \ion{Ti}{ii} & $-0.02\pm0.15$ & 13\\
      \ion{V}{i}  & $+0.05\pm0.10$ & 12\\
      \ion{Cr}{i} & $+0.03\pm0.27$ & 18\\
      \ion{Cr}{ii}& $+0.05\pm0.16$ & 3\\
      \hline
    \end{tabular}
    \hspace{2mm}
    \begin{tabular}{l c r}
      \hline \hline
      Elem. & [X/H] & $N$(X) \\
      \hline
      \ion{Mn}{i} & $+0.05\pm0.15$ & 5\\
      \ion{Co}{i} & $-0.13\pm0.10$ & 4\\
      \ion{Ni}{i} & $-0.10\pm0.10$ & 38\\
      \ion{Na}{i} & $-0.03\pm0.04$ & 2\\
      \ion{Al}{i} & $+0.01\pm0.10$ & 4\\
      \ion{Ba}{ii} & $+0.25\pm0.05$ & 3\\
      \ion{Ba}{ii}\tablefootmark{a} & $+0.13\pm0.09$ &  \\
      \ion{Li}{i}\tablefootmark{b} & $+1.55\pm0.38$ &  \\
      \hline
    \end{tabular}
    \end{center}
    \tablefoot{\tablefoottext{a}{See discussion in Sect.~\ref{sec:ExcitationIonization}.}
               \tablefoottext{b}{See discussion in Sect.~\ref{sec:Li}.}
              }
  \end{minipage}
\end{table}

\subsection{Effective temperature via line ratios}
\label{sec:LineRatios}

Comparing spectral lines with different temperature sensitivity with each other provides a valuable temperature diagnostic.
Particularly useful is comparing the EWs of spectral lines belonging to metallic species.
A line-ratio technique based on such a comparison was proposed by \cite{Sousa2007}, who also calibrated their method using $451$ FGK dwarf stars \citep{Sousa2010}.
The authors determine an empirical relation between effective temperature and the EW ratio for a set of $433$~pairs of spectral lines and
incorporated their results into the ``Teff\_LR Code'' code, which is an extension to the ARES code; both are freely available.

We used the relations published in \cite{Sousa2007,Sousa2010} and obtained a value of \mbox{\teff$=5513\pm111$~K} for \coa's effective temperature,
using 322 metallic lines and 22 independent line ratios. Although somewhat lower, this value is consistent with previous estimates (cf. Table~\ref{tab:CoRoT2specPars}).
As a cross-check, we determined the solar effective temperature using the same set of lines in the UVES solar spectrum and found the resulting value of $5784\pm152$~K to agree well with the literature \citep[][p.~341]{AllenEd4}.

\subsection{Synthetic spectra fitting via SME}
\label{sec:SME}
\label{sec:SMEteff}
\label{sec:SMElogg}

As an alternative to the modeling of line EWs, the stellar parameters can also be obtained by directly fitting the profile of spectral lines using synthetic spectra. This approach is implemented in the ``Spectroscopy Made Easy'' (SME) package \citep[version 2.1;][]{ValentiPiskunov1996}.
This interpolates on a Kurucz grid of stellar atmospheres and employs a VALD\footnote{see \url{http://www.astro.uu.se/~vald/php/vald.php}} \citep{Piskunov1995} line list to compute a synthesized spectrum for each set of stellar parameters. The observed spectrum is fitted by minimizing the residuals via a non-linear least-squares algorithm.

We used SME to determine the stellar parameters first in a global fit and second by fitting individual lines sensitive to \teff \ and $\log{g}$.
Currently, it is not feasible to compute a reliable error estimate for a global fit due to the large computational effort of calculating synthetic spectra. From the analysis of a set of 1040 FGK stars, however, 
\citet{ValentiFischer2005} derived typical errors of 44~K in \teff, 0.06~dex in $\log{g}$, and 0.03~dex in metallicity, which we adopt below. For the analysis of single line profiles, we usually fixed all parameters at their best-fit values and obtained the error by computing the 90\,\% confidence interval ($\Delta \chi^2=2.71$).

An approximation
of the effective temperature can be obtained by investigating the H$\alpha$ line profile.
The wings of the prominent H$\alpha$ line at a nominal wavelength of $6563$~\AA \
are sensitive to a wide range of effective temperatures of G- and F-type stars \citep[e.g.][]{Fuhrmann2004},
while remaining reasonably unaffected by the surface gravity, $\log{g}$, and the
metallicity.

However, active late-type stars are known to show strong contributions of chromospheric
emission in the Balmer lines, which can even extend into the wings of the line profiles \citep[e.g.,][]{Montes1997}. This can interfere with the determination of the effective temperature.
We independently analyzed the line wings of H$\alpha$ and H$\beta$.
Consistent results for the temperatures deduced from both Balmer lines indicate
that the wings of H$\alpha$ are not strongly affected by chromospheric
activity \citep{Fuhrmann2004, Koenig2005}.

A visual inspection of the symmetry of the H$\alpha$ line profile suggested that
the UVES pipeline, in this respect, provided
a superior result, so that we rely on the pipeline spectra during this analysis.
We note, however, that in any case a manual rectification of the spectrum is necessary.
Hence, this method is prone to considerable uncertainties.
We used SME to fit synthetic spectra to the observed H$\alpha$ and H$\beta$ line profiles excluding the line cores and found an effective temperature $T_\mathrm{eff}$ of $5510^{+90}_{-70}$~K and $5520^{+80}_{-90}$~K, respectively.
This is consistent with the result of $5450\pm120$~K derived from the analysis of the H$\alpha$ line observed with HARPS \citep{Bouchy2008}.
The good agreement between the Balmer line estimates indicates that the chromospheric contribution remains small.

Several pressure-broadened spectral lines can be used to determine the surface gravity of late-type stars. Examples of these lines are the \ion{Mg}{i}b triplet \citep{Fuhrmann1998}, the \ion{Na}{i}~D doublet, and the \ion{Ca}{i}~lines at 6122, 6162, and 6439~\AA \ \citep[e.g.,][]{Bruntt2010}. Because of the gap between the two detectors, the spectrum does not contain any \ion{Mg}{i}b lines. We therefore concentrated on the Na and Ca~lines. With SME, we iteratively fitted synthetic spectra to the three \ion{Ca}{i}~lines and the Na~D~line, leaving only $\log{g}$ as a free parameter. The resulting value from the \ion{Ca}{i} lines ($4.49\pm0.14$) was found to be consistent with the value derived from Na~D ($4.53\pm0.18$).

\subsection{The age of \coa \ as determined by \ion{Li}{i}}
\label{sec:Li}

The abundance of lithium is a valuable indicator of the stellar age. The element is depleted by lithium burning
primarily during the early phases of stellar evolution, when the existence of deep convection zones allows for
the interchange of material between the stellar interior and the surface \citep[e.g.,][]{Pinsonneault1994}.

\coa \ shows a strong Li~line at $\approx 6708$~\AA, for which we determined an EW of $139\pm1$~m\AA. We used SME to fit synthetic spectra with all remaining stellar parameters kept fixed and derived an abundance of $A_{\mathrm{Li}}=+2.6\pm0.3$. This value confirms the result of \cite{Gillon2010}, who found $A_{\mathrm{Li}}=+2.8$. According to \cite{SestitoRandich2005}, this a Li content is typically found in G-type stars of \teff$=5600$~K at an age between 100 and 250~Ma.

\begin{figure}[t]
  \includegraphics[width=0.5\textwidth]{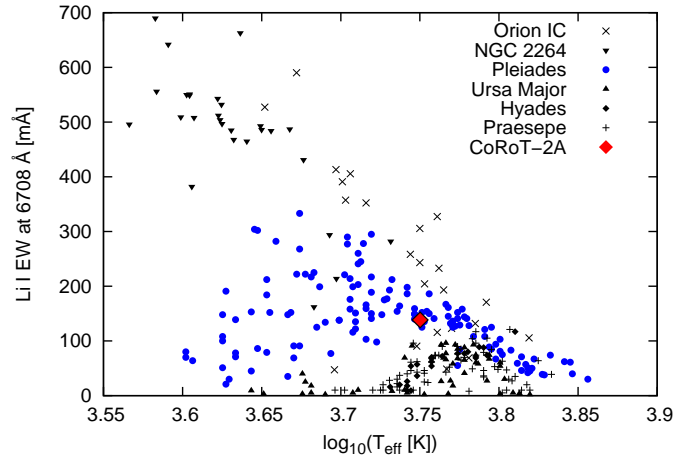}
  \caption{\ion{Li}{i} EW vs. effective temperature for \coa \ and a sample of open stellar clusters with different ages.
  \label{fig:LiCluster}}
\end{figure}

In Fig.~\ref{fig:LiCluster} we show effective temperature versus \ion{Li}{i} line EW for the
open stellar clusters
Orion~IC ($10$~Ma), NGC~2264 ($10$~Ma), Pleiades ($100$~Ma), Ursa Major ($300$~Ma), Hyades ($660$~Ma), and Praesepe ($660$~Ma)
\citep{King1993, Soderblom1999, Soderblom1993, Soderblom1993b, Soderblom1990, Soderblom1993c}, additionally,
the location of \coa \ is marked in the diagram.
The clusters are of different age, so that putting \coa \ in the context of the cluster properties
provides an indication of its age. 
The \ion{Li}{i} EW of \coa \ is best compatible with those
in the Pleiades, indicating an age of about $100$~Ma.
This finding is consistent with the numbers derived by \citet{GuillotHavel2011} from evolutionary modeling and
also the age estimates given by \citet{Gillon2010}, who derive an age between $30$ and $316$~Ma.

\subsection{Activity indicators}
\label{sec:ActivityIndicators}

\begin{figure}[t]
\includegraphics[width=0.5\textwidth]{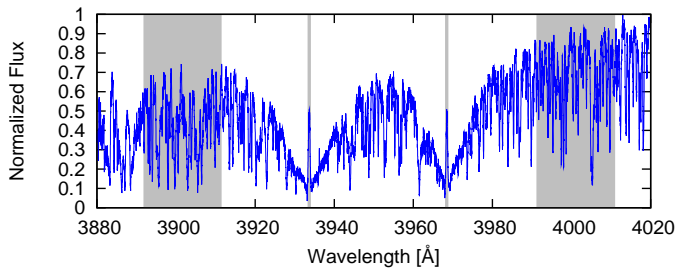}
\caption{Region of \coa's spectrum containing the \cahk \ lines.
Gray shades indicate the bands defined by
\citet{Melo2006} for the definition of $\log{R'_{\mathrm{HK}}}$. See text for details.
\label{fig:CaHK}}
\end{figure}

Active stars are known to show strong emission in the \cahk \ line cores at 3934.8 and 3969.7~\AA \ \citep[see][for a profound discussion]{Linsky1980}. This is also true for \coa; we show its \cahk \ lines in Fig.~\ref{fig:CaHK}.

In late-type stars the width, $W_0$, of the emission cores seen in the the \cahk \ lines is  mainly sensitive to the value of $\log g$, and, hence, to the mass and radius of the star, but insensitive to the effective temperature and metallicity. If calibrated appropriately, the width can be used as a rough estimator of the absolute luminosity of a star \citep{WilsonBappu1957}. We used the recent calibration of \citet{Pace2003} together with \coa's apparent magnitude corrected for interstellar extinction (see Sect.~\ref{sec:distanceToCorot2}) to obtain a distance estimate for the \co \ system. 
\citeauthor{Pace2003} find no significant effect of rotational and instrumental broadening within their sample, which comprises
stars with $v\sin{i}<14$~\kms. Neglecting broadening effects, we obtained a distance estimate of $190_{+60}^{-50}$~pc.
When both broadening mechanisms were taken into account by a quadratic correction to $W_0$ as described by \citeauthor{Pace2003}, the distance estimate reduced to $140_{-40}^{+50}$~pc, which
is lower but still compatible, given the errors.

\begin{figure*}[t]
\includegraphics[width=\textwidth]{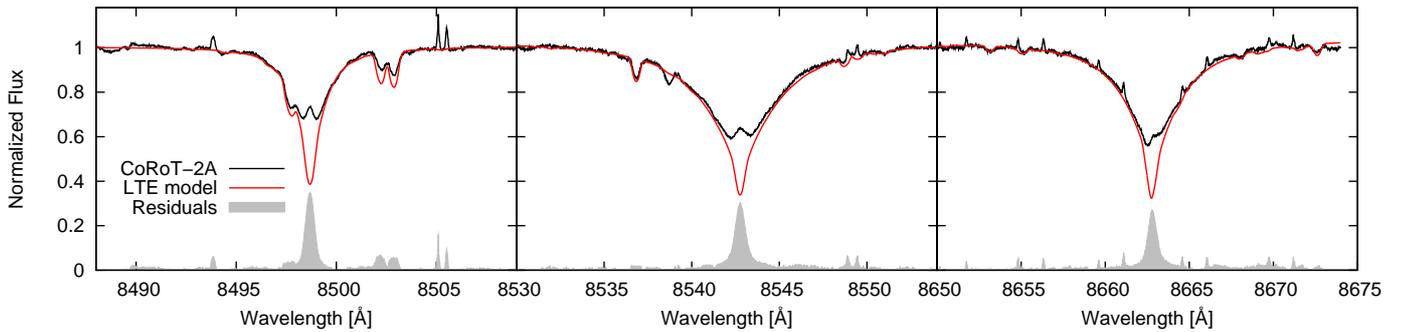}
\caption{\ion{Ca}{ii} infrared triplet in the spectrum of \coa \ (black), which is used
as activity diagnostic. The activity indicators are calculated relative to an LTE synthetic spectrum (red) based on the residuals (gray) as outlined in \citet{Busa2007}.
The small emission features result from sky emission (cf., Sect.~\ref{sec:Observations}).
\label{fig:CaIRT}}
\end{figure*}

The Mt. Wilson chromospheric flux index \citep[$S_{\mathrm{MW}}$;][]{Vaughan1978, Baliunas1998} is a popular measure of chromospheric activity. To estimate $S_{\mathrm{MW}}$ from our spectra, we used the calibration procedure for UVES spectra described by \citet{Melo2006}. 
These authors defined a proper index $S_{\mathrm{US}}$ and determined the following relation between their index and the Mt. Wilson index:
$$S_{\mathrm{US}} = 0.06111 \times S_{\mathrm{MW}} - 0.00341$$
\citep[see Eq.~1 in][]{Melo2006}. Inverting this relation, we obtained an $S_{\mathrm{MW}}$ index of $0.479$.
We proceeded by correcting our $S_{\mathrm{MW}}$ index for the color dependence \citep{Rutten1984} and subtracted the expected photospheric contribution to the flux in the line cores \citep{Noyes1984}.
Using a value of $0.854$ for the B-V color of \coa \ from the SIMBAD database\footnote{see \url{http://simbad.u-strasbg.fr/simbad/}}, we obtained an emission ratio of $\log{R'_{\mathrm{HK}}}=-4.458\pm0.051$, which agrees well with the results of \citet{Gillon2010} ($-4.471\pm0.0629$) and \citet{Knutson2010} ($-4.331$).

The central part of the \cahk \ emission core does not show the double-horned structure usually observed in the line cores of solar-like stars. We found no sign of self-reversal in the central part of the line core.
According to \citet{Ayres1979}, strong chromospheric heating through stellar activity leads to a decrease of the wavelength separation between the two K$_2$ peaks of the emission core, so that the central K$_3$ dip is easily obliterated by instrumental, macroscopic, and microscopic broadening effects. The lack of a detectable self-reversal is, therefore, an indicator of strong chromospheric heating itself. Because chromospheric heating causes an increase in the wavelength separation of the two K$_1$ minima that approximately counterbalances the mutual approach of the K$_2$ peaks, the Wilson-Bappu width, $W_0$, remains basically insensitive to stellar activity.

Chromospheric activity can also be measured in the \ion{Ca}{ii} infrared triplet (IRT) at 8498, 8542, and 8662~\AA \ \citep[see e.g.,][]{Andretta2005, Busa2007}. To correctly assess the chromospheric contribution to the line profile, the line forming process within the photosphere must be accurately modeled. \cite{Andretta2005} proposed the activity indicator $R_{\mathrm{IRT}}$ defined as the difference between the central line depths (or central depressions) of the observed spectrum and a rotationally-broadened NLTE model. We calculated the central depression, $R_{\mathrm{IRT}}$, and $\Delta W_{\mathrm{IRT}}$, i.e., the EW of the residual line profile and the corresponding error estimates following the approach detailed in \cite{Busa2007}.
In our calculations, we used LTE models synthesized via SME, which,
according to \citeauthor{Andretta2005}, suffice to approximate the photospheric spectrum for main-sequence stars of solar metallicity. 
The parameters thus obtained are summarized in Table~\ref{tab:CoRoT2CaIRT}. The \coa \ Ca~IRT along with our synthetic templates are shown in Fig.~\ref{fig:CaIRT}, which also indicates the residuals.
\citeauthor{Busa2007} found a relationship between the $\log{R'_\mathrm{\mathrm{HK}}}$ activity index and $R_{\mathrm{IRT}}$, however, with a large scatter. Given our value of $\log{R'_\mathrm{HK}}=4.46$, the relation predicts a value of $0.25$ for $R_{\mathrm{IRT}}$, which reasonably agrees with our results.

\begin{table}[t!]
  \begin{minipage}[h]{0.5\textwidth}
    \renewcommand{\footnoterule}{}
    \begin{center}
      \begin{tabular}{c c c c}
      \hline \hline
      $\lambda$ [\AA] & $CD_\mathrm{obs}$ & $R_\mathrm{IRT}$ & $\Delta W_\mathrm{IRT}$ [m\AA]  \\
      \hline
      $8498$ & $0.266\pm0.009$ & $0.342\pm0.009$ & $237.6\pm1.6$ \\
      $8542$ & $0.361\pm0.006$ & $0.297\pm0.006$ & $63.1\pm0.6$ \\
      $8662$ & $0.413\pm0.014$ & $0.256\pm0.014$ & $171.4\pm2.1$ \\
      \hline
      \end{tabular}
    \end{center}
    \caption{Central depression ($CD$), $R_\mathrm{IRT}$ and residual equivalent width $\Delta W_\mathrm{IRT}$ for the lines of the \ion{Ca}{ii}~infrared triplet in \coa. \label{tab:CoRoT2CaIRT}}
  \end{minipage}
\end{table}

\subsection{Interstellar absorption features}
\label{sec:interstAbsDist}

\begin{figure}[t]
\includegraphics[width=0.5\textwidth]{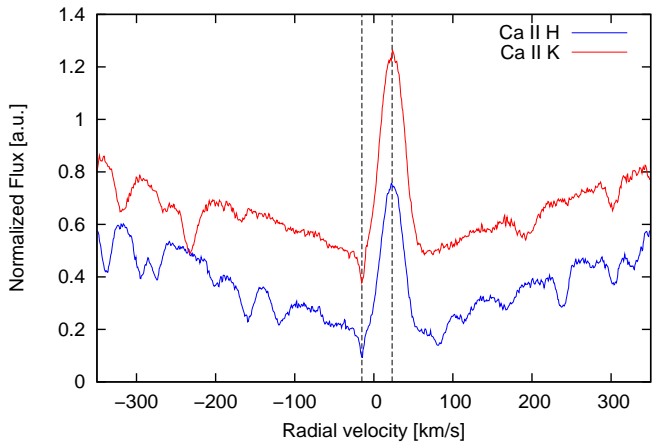}
\caption{Emission line cores of the \cahk \ lines clearly showing the H$_4$ and K$_4$ absorption features. The spectrum of \coa \ is shifted in radial velocity by $23.2$~\kms, while the absorption features show a barycentric velocity shift of $-15$~\kms.
\label{fig:the_wind_plot}}
\end{figure}

Analyzing the emission line cores of the \cahk \ complex, we found narrow K$_4$ \citep[see e.g.,][for designation]{Reimers1977} and H$_4$ absorption features close to both central chromospheric emission peaks (Fig.~\ref{fig:the_wind_plot}). We therefore attribute the features to \ion{Ca}{ii} absorption and determine a barycentric velocity of $-15$~\kms, corresponding to a relative velocity of about $+40$~\kms \ with respect to \coa.

Separately considering our $24$ UVES spectra observed within 6~hours and the archived UVES spectrum
taken about three years earlier in 2007,
we found no intrinsic variability of this feature.
The EW of the \ion{Ca}{ii}~K$_4$ absorption feature is
65~m\AA, slightly higher than the corresponding EW of the \ion{Ca}{ii}~H$_4$ feature with
53~m\AA. These blue shifted absorption features of ionized material may be caused by wind absorption as observed in giant stars \citep[e.g.,][]{Reimers1977} or, alternatively, by the interstellar medium.

While \cahkabs \ features caused by wind absorption are observed in giant stars rather than dwarfs such as \coa,
the narrowness of the features is surprising if they are attributed to interstellar absorption.
Often, multiple interstellar clouds are found in the line of sight,
giving rise to a more diffuse absorption
feature. Indeed, the width of the \cahkabs \ features is comparable to that of telluric lines in our spectrum.
However, considering the temporal stability,
we argue in favor of interstellar absorption as the origin of the \cahkabs \ features.

The Sun is situated within a region pervaded by hot, ionized plasma of low density known as the Local Cavity that reaches a radius of about $100$~pc \citep[e.g.,][]{CoxReynolds1987, Sfeir1999}. This cavity contains many diffuse clouds with a complex velocity structure, which typically give rise to \ion{Ca}{ii}
absorption features with EWs of a few m\AA \ \citep{LallementBertin1992}.
The density distribution of \ion{Ca}{ii} is fairly uniform in the interstellar medium beyond the Local Cavity, so that the EW of interstellar \cahk \ absorption can be used as a distance indicator
for sufficiently distant objects \citep{Megier2005, Megier2009, Welsh2010}.
Using Eqs.~1 and 2 in \citet{Megier2005}, we translated the EWs of both \ion{Ca}{ii} absorption features into distance estimates of
$274$~pc for the \ion{Ca}{ii}~K$_4$ feature and $256$~pc for the \ion{Ca}{ii}~H$_4$ feature.

\begin{figure}[t]
\includegraphics[height=0.5\textwidth, angle=-90]{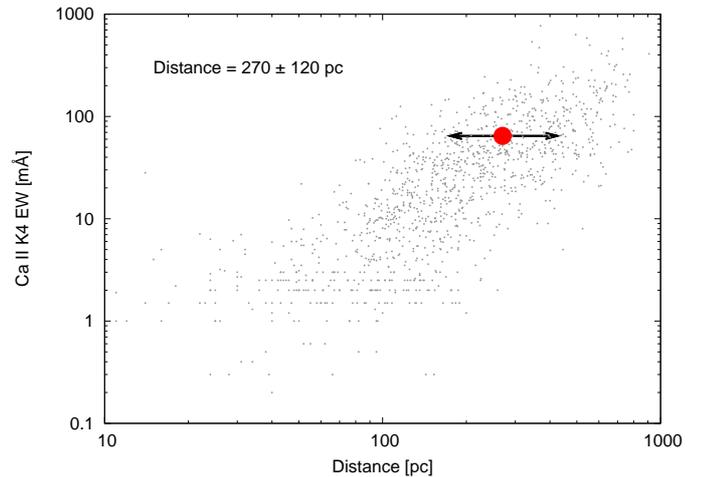}
\caption{Equivalent width of the interstellar \ion{Ca}{ii}~K$_4$ absorption line versus distance
for stars within 800~pc \citep[catalog data compiled by][]{Welsh2010}. The measured \ion{Ca}{ii}~K$_4$ absorption
of \coa \ is consistent with a distance of roughly 300~pc.
\label{fig:welsh}}
\end{figure}

We note that our spectrum of \coa \ also shows two strong, nearly saturated \ion{Na}{i} absorption lines with the same radial velocity shift within the blue wing of the \ion{Na}{i} D~doublet. These broad absorption features have EWs of $270$~m\AA \ (D$_2$) and $230$~m\AA \ (D$_1$), which is again compatible with interstellar absorption and a distance of $340\pm180$~pc and $320\pm140$~pc, respectively \citep{Welsh2010}.
Additional interstellar absorption lines of \ion{K}{i} at $7698.974$~\AA \ or molecules such as CH could not be detected.

In summary, we estimate
a distance of $270\pm 120$~pc for \coa, based on interstellar absorption; the
error was estimated from the standard deviation of the distances of the stars with K$_4$ EWs in a $\pm 5$~m\AA \ band around our measured EW (cf., Fig.~\ref{fig:welsh}).

\subsection{Evidence for a gravitationally bound companion}

\coa \ has a close neighbor, \tmass, about $4\arcsec$ in southeast direction. \citet{Alonso2008} found that
the color of this visual companion is consistent with a late-K or early-M type star located at the same distance as \coa.
During five of our 24 UVES observations this nearby neighbor was placed inside the slit along with \coa, which we used to obtain a low SNR spectrum of the companion. The separation of the two objects allowed us to separately extract the spectra using the UVES pipeline. 

Because the companion is $\approx 3.5$~mag fainter than \coa \ in the visual band, the resulting spectrum has an SNR of no more than $10-20$ depending on wavelength. It is dominated by absorption lines from neutral and singly-ionized metals. In particular, we find strong absorption in \ion{Ca}{i} and \ion{Mg}{ii}, whereas the \ion{Ca}{ii} lines are comparably weak. Furthermore, we find a relatively weak H$\alpha$ line and distinct edges caused by titanium oxide (TiO) absorption.

The TiO bands are a valuable indicator of stellar effective temperature if the metallicity is known, while they are less sensitive to surface gravity \citep{MiloneBarbuy1994}.
To obtain an estimate of the effective temperature, we compared the coadded companion spectrum to synthetic spectra calculated with SPECTRUM\footnote{see \url{http://www1.appstate.edu/dept/physics/spectrum}} \citep{GrayCorbally1994} using line lists containing the TiO and ZrO lines compiled by \cite{Plez1998}\footnote{see \url{http://www.graal.univ-montp2.fr/hosted/plez}}.

\begin{figure}[t]
  \includegraphics[width=0.5\textwidth]{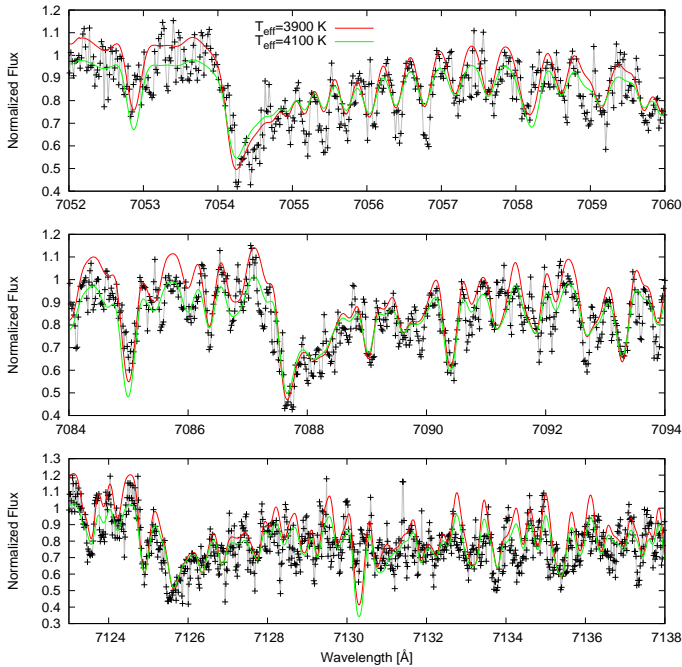}
  \caption{Part of the companion spectrum showing the temperature-sensitive TiO band at $7050 - 7200$~\AA.
  The red/green lines are synthetic spectra for \teff$=3900$~K and $4100$~K, respectively, computed for solar metallicity and $\log{g}=4.9$.
  \label{fig:TiObandhead}}
\end{figure}

Assuming solar metallicity, we set up a Markov-Chain Monte-Carlo (MCMC) framework to find an estimate of the effective temperature and the associated error. We used uniform priors on the stellar parameters and allowed for an additional normalization constant accounting for inadequacies during blaze correction and continuum normalization. We focused on the strongest TiO band with its bandhead at $7054$~\AA \ and analyzed the three absorption edges individually.
The resulting three 95\,\% credibility intervals for \teff \ consistently yield an effective temperature between 3900 and 4100~K (see Fig.~\ref{fig:TiObandhead}). The surface gravity was found to be $\log{g}=4.9\pm0.1$, which is very close to our imposed upper bound of $\log{g}=5$ and does not seem to be well constrained.

We tried to use SME to fit synthetic spectra to several spectral lines known to be sensitive to the stellar parameters. Our efforts were, however, strongly hampered by the low SNR of the spectral data owing to the faintness of the companion.
From the analysis of the \ion{Na}{i} doublet at 5890~\AA, the \ion{Ca}{i} lines at $6122$, $6162$, and $6439$~\AA, and a set of 137~single Fe~lines, we found the stellar spectrum to be best described by effective temperatures around 4000~K and a surface gravity between $4.6$ and $4.9$.
Unfortunately, the quality of the spectrum made an analysis of the H$\alpha$ line impossible.
The set of \ion{Fe}{i} lines was also used to quantify the effect of rotational line broadening.
Neglecting additional line broadening effects, we obtained an upper limit of 10~\kms \ on $v\sin{i}$ at a confidence level of $90$\,\%.

These findings are compatible with the companion being a K9~star \citep[][p.~388, Table~15.7]{AllenEd4}. Our results are in line with those of \citet{Alonso2008} and \citet{Gillon2010}, who found the photometric magnitudes measured in optical (Exodat), near-infrared (2MASS), and infrared (\textit{Spitzer}) filter bands to be consistent with a late-K or early-M type companion star.

To find the radial velocity of the visual companion, we cross-correlated our five companion spectra with a template spectrum corresponding to our best-fit stellar parameters. The radial velocity was estimated independently in our five companion spectra, and corrected for the wavelength drift visible in the telluric lines. The resulting average radial velocity amounts to $23.9\pm0.4$~\kms, a value close to \coa's radial velocity of $23.245\pm0.010$~\kms, which we determined accordingly. We note that it is also
independent of the details of the chosen spectral model.
Given the apparent distance in the sky of $4\arcsec$ and a distance of about 270~pc the projected distance between \coa \ and the companion amounts to about 1100~AU. Because \coa \ is basically solar-like in mass, Kepler's third law yields a lower bound for the orbital period of $40\,000$~a, which gives an orbital velocity of up to 0.9~\kms. Thus, the radial velocity found for the visual companion of \coa \ is compatible with the hypothesis of a gravitationally bound companion.

To check for relative sky motions of \coa \ and \tmass, we inspected the photographic data available from the Digitized Sky Survey.
We checked the digitized plates of the Palomar Observatory Sky Surveys from 1951, 1983, and 1991 and the HST Guide Star Catalogue from 1980, but found no indications for a relative transversal motion. This finding corroborates the hypothesis that \tmass \ and \coa \ are gravitationally bound and form a wide binary system.

The parameters derived for the companion are summarized in Table~\ref{tab:CoRoT2BspecPars}.

\begin{table}[t!]
  \begin{minipage}[h]{0.5\textwidth}
    \renewcommand{\footnoterule}{}
    \caption{Stellar parameters of \tmass. \label{tab:CoRoT2BspecPars}}
    \begin{center}
      \begin{tabular}{l l l}
      \hline \hline
      Parameter & Value & Lines \\
      \hline
      \teff\ [K] & $[3900,4100]$ (95\,\% cred.) & TiO $7050-7200$~\AA \\
      $v\sin{i}$ [\kms] & $<10$ (90\,\% CL) & 137 \ion{Fe}{i} lines\\
      RV [\kms] & $23.9\pm0.4$ & \\
      \hline
    \end{tabular}
    \end{center}
  \end{minipage}
\end{table}

\section{Analysis of the \chan \ X-ray data}
\label{sec:AnalysisofXraydata}

\subsection{Observations}

\coa \ was observed by \chan \ using the ACIS-S detector on June~24, 2010 for about 15~ks (Obs.-ID 10989). In the reduction and analysis process, we used the standard software package CIAO in version 4.2. To obtain the best possible timing, the tool \texttt{axbary} was used to apply a barycentric correction to the photon arrival times.

\subsection{Detection and spectral analysis}

\begin{figure}[t]
  \includegraphics[angle=-90, width=0.45\textwidth]{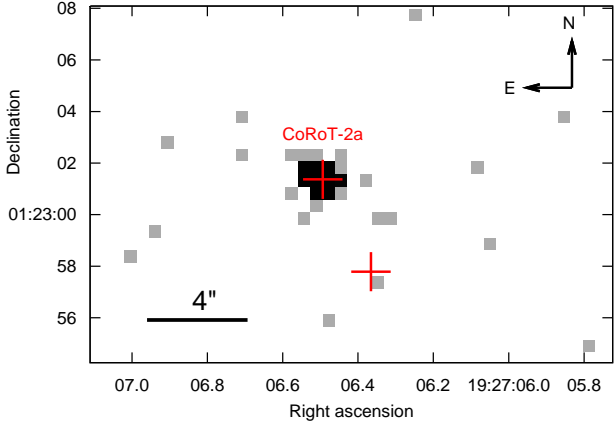}
  \caption{X-ray image ($0.5\arcsec$/bin) of the surrounding of \coa \ for energies above $0.3$~keV. The nominal 2MASS positions of \coa \ and \tmass \ are marked with red crosses.}
  \label{fig:CoRoT2ds9}
\end{figure}

In a first step, we screened the X-ray image for photons in the $0.3-4$~keV energy band. This step
reduces the background contamination and focuses our analysis on the energy band, in which stellar coronal
emission is expected to dominate. We show parts of the resulting X-ray image in Fig.~\ref{fig:CoRoT2ds9}.

In a second step, we counted all photons within a $2\arcsec$~radius circular region centered on the nominal position of \coa. In this region, we found $87$~photons with an
expected background contribution of $\approx$\,3~photons,
deduced from nearby source-free regions.

Finally, we carried out a spectral analysis of the source photons.
Using XSPEC~v12.5 \citet{Arnaud1996},
we fitted the ACIS spectrum with an absorbed, thermal APEC \citep[e.g.,][]{Smith2001} model.
Because the abundances are not well constrained by the fit,
we fixed them at their solar values for the rest of the analysis, which is in accordance with our optical estimates (cf. Sect~\ref{sec:ExcitationIonization}).
For the depth of the absorbing column, the fit provides a value of \mbox{$\approx 10^{21}$~cm$^{-2}$}, which is well compatible with a canonical density of \mbox{$1$~particle per cm$^3$} for the interstellar medium
and a distance of $\approx 270$~pc (see Sect.~\ref{sec:interstAbsDist}) for \coa.

\begin{figure}[t]
  \includegraphics[angle=-90, width=0.49\textwidth]{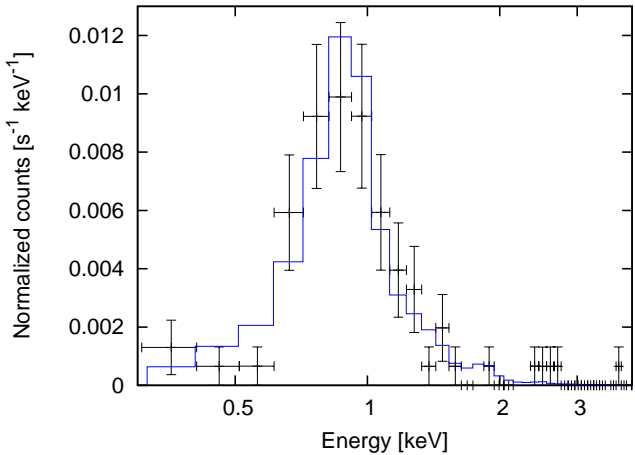}
  \caption{ACIS-S spectrum of CoRoT-2a with applied spectral model.}
  \label{fig:CoRoT2spec}
\end{figure}

Figure~\ref{fig:CoRoT2spec} shows the X-ray spectrum and our best-fit model; the fit results are
summarized in Table~\ref{tab:CoRoT2spec}.
From our best-fit model, we obtain an X-ray luminosity of $L_\mathrm{X} = 1.9\times 10^{29}$~\ergs \ in the $0.3-4$~keV band,
corresponding to an activity level of $\log_{10}{L_\mathrm{X}/L_{\mathrm{bol}}} \approx -4.2$, which indicates that \coa \ is an active star also by X-ray standards.

\begin{table}[t!]
  \begin{minipage}[h]{0.5\textwidth}
    \renewcommand{\footnoterule}{}
    \caption{Spectral-fit results for CoRoT-2a derived from \chan \ ACIS-S data. \label{tab:CoRoT2spec}}
    \begin{center}
      \begin{tabular}{l l}
      \hline \hline
      Parameter  & Value (90\,\% conf.) \\
      \hline
      $N_\mathrm{H}$ [$10^{22}$~cm$^{-2}$] & $<0.2$ \\
      $T$ [keV] & 0.74 (0.57-0.83) \\
      $T$ [$10^6$~K] & 8.6 (6.6-9.6) \\
      $f_\mathrm{X}$ [$10^{-14}$~\ergcms] & 2.1 (0.8-2.9) \\
      $L_\mathrm{X}$ (at $270$~pc) [$10^{29}$~\ergs] & 1.9 (0.7-2.5)\tablefootmark{a} \\
      \hline
      \end{tabular}
    \end{center}
    \tablefoot{\tablefoottext{a}{An error of $\pm120$~pc on the distance estimate translates into a larger conf. interval for $L_\mathrm{X}$ of $(0.2-5.2)\times 10^{29}$~\ergs.}}
  \end{minipage}
\end{table}

\subsection{X-ray light curve and transit}

\begin{figure}[t]
  \includegraphics[angle=-90, width=0.49\textwidth]{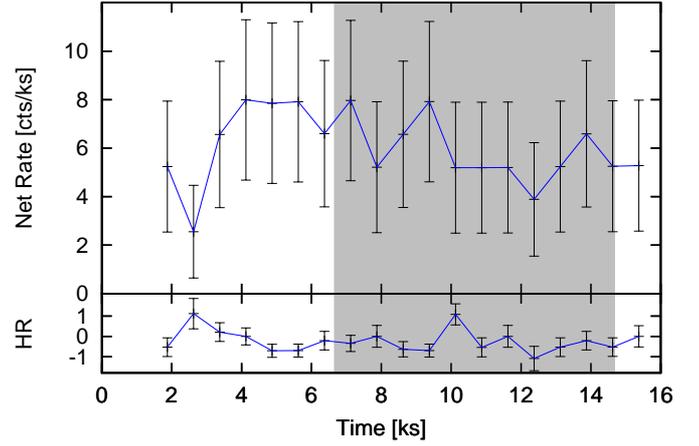}
  \caption{Light curve of CoRoT-2a in the $0.3-4$~keV band and hardness ratio $(H-S)/(H+S)$ for $0.3-1$ and $1-4$~keV bands (750~s binning). The shaded area corresponds to the transit in visual wavelengths.}
  \label{fig:CoRoT2lc}
\end{figure}

To investigate the X-ray variability of \coa, we constructed background-subtracted light curves with various binnings.
A barycentric time-correction was applied to all light curves to obtain time stamps, which can be easily reconciled with planetary ephemerides
given in the literature.
Figure~\ref{fig:CoRoT2lc} presents the light curve of \coa, which shows no indications of strong short-term variability like flares, and, therefore,
we argue in favor of quiescent emission.

The ephemerides of \cob \ were derived by \citet{Alonso2008} using the CoRoT data.
The transit duration is $8208$~s,
during which the relative flux deficit in the optical
reaches $3$\,\%.
Our \chan \ observation completely covers
one planetary transit \citep[epoch 651 with respect to the ephemerides from][]{Alonso2008}. The ingress
begins $6538$~s after the start of the observation and the egress is finished shortly before the observation ends (cf., Fig.~\ref{fig:CoRoT2lc}).

Our analysis showed that the source count rate, if anything, increased by $17$\,\% during the eclipse.
Similarly, the hardness ratio $HR = (H - S )/(H + S )$ with $S = 0.3-1$~keV and $H = 1-4$~keV
(lower panel in Fig.~\ref{fig:CoRoT2lc}) remains unaffected.
On the one hand, given $82$~source counts in 15~ks,
detecting a $3$\,\% drop in brightness as observed in the optical seems to be out of reach in X-rays.
On the other hand, the sources of X-ray emission are
believed to be distributed much more inhomogeneously across the stellar surface than those of optical light.
We conclude that either the planet did not eclipse a strong concentration of X-ray emitting material in this particular case, the emission is distributed homogeneously, or concentrated at higher latitudes avoided by the planetary disk.

\subsection{The companion in X-rays}

To check whether \coa's potentially physical companion, \tmass, is an X-ray source, we collected the photons within a circle of $1\arcsec$ radius centered on the star's
2MASS position. According to our modeling, this region contains $93$\,\% of the \chan \ point spread function (PSF) at $1$~keV.
A single photon with an energy of $1.1$~keV was detected in this region. Because $99$\,\% of \coa's PSF are confined to a distance
of $4\arcsec$ and less from \coa, the detected photon is unlikely to stem from that source.
From nearby source-free regions, we estimated the rate of background-counts with energies of $1.1 \pm 0.1$~keV, where the $0.1$~keV
range accounts for \chan's \ energy resolution, to be $2\times 10^{-4}$~cts\,s$^{-1}$ within the
encircled region centered on the companion.

The detected photon may, consequently, be associated with an X-ray source at the companion's
position.
To derive an upper limit on the X-ray flux of the companion, we determined the count rate yielding one or less detected photons with a probability of $95$\,\%. Consulting Poisson statistics, the limiting count rate amounts to $0.36$~cts in 15~ks or $2.4\times 10^{-5}$~cts\,s$^{-1}$.
Assuming that the source has a $1$~keV thermal spectrum, a distance of $270$~pc (cf. Sect.~\ref{sec:interstAbsDist}), and neglecting absorption, we used
WebPIMMS\footnote{\url{http://heasarc.gsfc.nasa.gov/Tools/w3pimms.html}} to convert the count rate into
an upper limit of $L_\mathrm{X}<9\times10^{26}$~\ergs for the companion's X-ray luminosity.

\section{Discussion}
\label{sec:Dicussion}

We presented new X-ray and optical data of the active planet host-star \coa.
Below, we discuss the impact of our findings on our understanding of the \co \ system.

\subsection{\coa's atmosphere and activity}

We studied the photosphere of \coa \ applying different techniques of spectroscopic analysis to determine the
stellar effective temperature, surface gravity, metallicity, and microturbulence velocity.
First, we measured the EW of 238~\ion{Fe}{i}~and~\ion{}{ii} lines and determined the entire set of parameters by imposing
excitation and ionization balance. Second, we redetermined the parameters by directly fitting several
sensitive spectral lines and found consistent results, which are, moreover, 
in line with previously published values \citep{AmmlervonEiff2009, Bouchy2008}.
With an effective temperature of $5598\pm34$~K and a surface gravity of $\log{g}=4.47\pm0,14$, \coa \ is
a star of spectral type G6--G7 \citep[][p. 151, Table 7.5]{AllenEd4} with slightly increased metallicity compared to the Sun. We compared the stellar parameters to those predicted by theoretical isochrone calculations for pre-main and main-sequence stars \citep{Siess2000}. Assuming that \coa \ is close to the zero-age main sequence, these models favor a spectral type of G7 given the observed values of effective temperature and metallicity.

The fact that \coa \ is a highly active star became first obvious in the photometry observed by the CoRoT observatory.
The light curve shows pronounced rotational variability caused by active regions that cover a substantial fraction of the stellar
photosphere \citep[e.g.,][]{Wolter2009, Czesla2009, Huber2009, Huber2010}.
This high level of activity, mainly diagnosed by photospheric spots, is expected to be also
detectable in chromospheric lines excited by enhanced chromospheric heating.

Indeed, \coa \ shows strong chromospheric emission-line cores in its \cahk~lines as well as in the Ca~IRT lines. We quantified
the strength of the emission by determining the Mt. Wilson S-index and a $\log{R'_{\mathrm{HK}}}$ value of $4.458\pm0.051$. Our results agree well with
those reported by \citet{Knutson2010} and place \coa \ among the most active known planet host-stars.
Furthermore, we studied the Ca~IRT and derived a value of $0.298\pm0.006$ for the $R_\mathrm{IRT}$ index, which again demonstrates that \coa \ is a highly active star.

The presence of starspots and strong chromospheric heating suggests that coronal heating is substantial as well.
Indeed, our 15~ks \chan \ observation yields a clear detection of coronal X-ray emission characterized by a thermal spectrum with a temperature of $1$~keV. Combining \coa's X-ray luminosity of $1.9\times 10^{29}$~\ergs \ with 
its spectral type of G7,
we derived an activity level of \mbox{$\log{L_{\mathrm{X}}/L_{\mathrm{bol}}} \approx -4.2$}, showing that \coa \ is a very active star also by X-ray standards.
Although optical studies \citep{Lanza2009, Huber2010} suggested a large inhomogeneity in the distribution of active regions, which is very likely also true for the
distribution of X-ray emission across the stellar disk, 
an X-ray transit could neither be detected in the X-ray count rate nor in the hardness ratio.
This indicates that either no prominent source of X-ray emission was occulted during this particular transit or that the emission is
distributed too homogeneously to cause an X-ray transit detectable with \chan.
In any case, we emphasize that our \chan \ snapshot covers no more than $4$\,\% of \coa's rotation period and a virtually negligible fraction of
the optically observed ``beating pattern'' \citep[e.g.,][]{Alonso2008} with a period of $\approx 50$~d, so that
it remains insufficient to obtain a representative picture of \coa's corona.

\subsection{The age of \co}

One of the key quantities needed to understand the evolution not only of the \co \ system but of all planetary
systems is their age. Based on our analysis, we applied several techniques to estimate the age of \coa.

From the EW of the lithium line at $6708$~\AA \ in the spectrum of \coa, we inferred an age
comparable with that of the Pleiades, i.e., $\approx 100$~Ma. Furthermore, we derived a Li
abundance of $A_\mathrm{Li}=+2.6$~dex, which suggests an age between 100 and 250~Ma.
Applying the relation provided by \citet[][Eq.~1]{Donahue1998},
we used the strength of the \cahk \ emission-line cores measured by the $\log{R'_{\mathrm{HK}}}$ index to estimate
a ``chromospheric age'' of $670^{+200}_{-280}$~Ma for \coa. 
The coronal activity provides another age estimate. Using the relation between X-ray luminosity and age for late-F to early-M dwarfs presented in \citet{SanzF2010}, we calculated an age of $230_{-40}^{+200}$~Ma.
An additional estimate can be
obtained from gyrochronology. Using the relation presented by \citet{Barnes2007},
we determined an age of $76\pm7$~Ma for \coa. However, \citeauthor{Barnes2007} note that gyrochronology tends to underestimate the stellar age if $($B-V$)>0.6$,
which is true for \coa, owing to the sparseness of the open cluster sample used for calibration in case of blue stars.

\citet{GuillotHavel2011} modeled the evolution of the star, \coa, and its planet simultaneously and
found two classes of solutions reproducing the observed properties of the \co \ system:
first, a solution in which \coa \ is a very young star with an age of $30$~to~$40$~Ma and second, a solution
with a more evolved main-sequence host-star with an age of $130$~to~$500$~Ma.
None of the above age indicators, not even the gyrochronological age estimate, favors the solution with a
very young host-star, rendering this class of evolutionary scenarios
found by \citet{GuillotHavel2011} unlikely.

In summary, we conclude that combining our outcomes with published results both observational and theoretical favors an age between 100 and 300~Ma for \coa. This suggests that \coa \ is a young main-sequence star that has already left the zero-age main sequence, which for a G7 star of solar mass is situated at an age of 30~Ma \citep{Siess2000}.

\subsection{Distance to \coa}
\label{sec:distanceToCorot2}

Photometric colors are often used for a rough spectral classification. The magnitudes provided by SIMBAD yield a B-V color index of $0.854$~mag for \coa. Comparing this value with the color index expected for a G7 star of age $200$~Ma with slightly increased metallicity (Z=$0.01$) \citep{Siess2000}, we calculated a color excess of \mbox{E$($B-V$)=0.15$~mag}. Thus, \coa \ appears redder than expected; we attribute this to interstellar extinction.

Combining the B-V color excess with the relation given by \citet{Bohlin1978}
$$
  <N(\ion{H}{i}+\ion{H}{ii})/\mbox{E(B-V)}> = 5.8 \times 10^{21}\;\mbox{cm$^{-2}$ mag$^{-1}$}\, ,
$$
we obtain a column density of $9\times 10^{20}$~cm$^{-2}$, which is consistent with  the upper limit of $2\times 10^{21}$~cm$^{-2}$ derived from our X-ray observations.
Converting the EWs of the interstellar Na~D absorption lines into a column density of \mbox{$8\times10^{12}$~cm$^{-2}$} via the line-ratio method \citep{Stromgren1948}, we obtained another consistent estimate of $\approx 10^{21}~\textrm{cm}^{-2}$ for the hydrogen column density \citep{Ferlet1985}. 
Assuming a density
of one particle per cm$^{3}$ for the interstellar medium, the hydrogen column density inferred from the color excess directly translates into a distance estimate
of $290$~pc, which is consistent with our previous estimate of $270$~pc.

According to our spectroscopic analysis, \coa \ can be classified as a G7-type dwarf and, according to the evolutionary model, should have an absolute visual brightness of $5.1$~mag.
Assuming a value of $R=3.1$ \citep{Schultz1975} for the ratio of total visual extinction, A$_V$, and \mbox{E(B-V)}, we derive \mbox{A$_V = 0.48$~mag} for \cocapa.
Combining this with the apparent visual brightness of $12.57$~mag, we derived an extinction-corrected spectroscopic parallax of $250$~pc.

In Sect.~\ref{sec:ActivityIndicators} we estimated the distance to the \co \ system from the Wilson-Bappu width of the \cahk \ emission line cores and obtained $140_{-40}^{+50}$ to $190_{-50}^{+60}$~pc depending on the details of the calibration. We further determined the distance from the presence of interstellar Ca and Na absorption features in the spectrum and obtained $270\pm120$~pc based on the interstellar absorption column (Sect.~\ref{sec:interstAbsDist}).
In summary, we argue in favor of a distance of $\approx 270$~pc as the most
likely value.

\subsection{X-rays eroding \cob}

Because \cob \ orbits its host star at a distance of only $0.03$~AU, it is immersed in an enormous high-energy
radiation field.
According to \coa's X-ray luminosity,
the X-ray flux at the distance of \cob's orbit amounts to \mbox{$8.5\times 10^{4}$~\ergcms }, which is five orders of magnitude larger than the solar X-ray flux received by Earth.
This amount of ionizing radiation can have a significant
influence on the structure and evolution of the planetary atmosphere.
\citet{Schneider1998} found that the extent of the atmospheres of hot Jovian exoplanets
can exceed the Roche lobe, leading to evaporation of planetary material by an interaction with the stellar wind.
Indeed, extended atmospheres of extrasolar planets were found 
for HD~209458~b \citep{Vidal-Madjar2003,Linsky2010} and HD~189733b \citep{LecavelierDesEtangs2010}.

\citet{SanzF2010} analyzed a sample of
planetary systems. The authors come to the conclusion that erosion triggered by stellar
high-energy illumination has a detectable influence
on the observed mass distribution of exoplanets.
This gives rise to an ``erosion line'', below which \citet{SanzF2010} \textbf{find} the large majority of the planets
in their sample.
To estimate the mass loss induced by the X-ray and extreme-UV (EUV) irradiation, we used Eq.~2 from \citet{SanzF2011}, viz.
$$\dot{M} = \frac{3 F_\mathrm{XUV}}{4G\rho} \; ,$$
where $F_\mathrm{XUV}$ is the sum of the stellar X-ray and EUV flux at the planetary orbit, $G$ is the gravitational constant, and $\rho$ is the density of
the planet (all in cgs units).
Because there are no EUV data of \coa \ available, we use Eq.~3 from \citet{SanzF2011}, which provides a relation between X-ray and EUV luminosity calibrated with their sample of objects, to obtain an estimate of $4.3\times10^5$~\ergcms \ for the expected EUV flux at the distance of \cob.
Substituting the parameters for \cob \ \citep[][Table~1]{Alonso2008}, we obtained a mass-loss rate of
$4.5\times 10^{12}$~g\,s$^{-1}$ or $7.3\times 10^{-2}$~M$_\mathrm{J}$\,Ga$^{-1}$ for the planet.
Given the uncertainties, this value remains a coarse estimate, but
places \cob \ clearly above the erosion line, which,
according to \citeauthor{SanzF2010}, may be explained by the youth of the system.
We note, however, that by using different assumptions for the extent of \cob's atmosphere \citep[][Eq.~1]{SanzF2010},
the mass-loss rate may be increased by up to one order of magnitude. Moreover, the effects leading to planetary
mass loss are not yet well understood.

\subsection{The companion - a puzzling genesis}

An important consequence of our analysis is that the \co \ system may
extend far beyond the planetary orbit. \coa's visual companion, \tmass, may actually also be a physical companion, forming
a wide binary pair with \coa.

We obtained and analyzed the first low-SNR UVES spectrum of the companion.
\citet{Alonso2008} already noticed that the companion may be a late-K or early-M-type star at about
the same distance as \coa. We measured the companion's radial velocity and found a value of $23.9\pm0.4$~\kms,
which is close to \coa's radial velocity.
By modeling the TiO bands present in the spectrum,
we determined an effective temperature between 3900 and 4100~K. Wide lines of \ion{Ca}{i} were used to infer a surface gravity of $\log{g}=4.74$.
Consulting the evolutionary tracks of \citet{Siess2000} at an estimated age of $200$~Ma, we find that
the companion is likely to be a star of spectral type K9, which is gravitationally bound to \coa.
This would make \co \ one of about 40~known binary systems harboring an exoplanet \citep{Mugrauer2009}.

If this hypothesis withstands further observational tests, it would challenge our understanding of the
\co \ system, in particular, the age of the system.
From the \chan \ data we derived an upper limit of \mbox{$9 \times 10^{26}$}~\ergs \ for the X-ray luminosity of the companion, and
the \citeauthor{Siess2000} evolutionary tracks suggest an absolute bolometric luminosity of $6.8$~mag. Combining these numbers, we obtain \mbox{$\log{L_{\mathrm{X}}/L_{\mathrm{bol}}} < -5.8$}, making the companion a star much less active than \coa.
From our spectral analysis, we concluded that \coa \ has an age between $100$ and $300$~Ma. 
Assuming that the companion is physically bound and has the same age as \coa, we would expect an X-ray flux significantly higher than observed in our \chan \ pointing. From the study of X-ray emission of members of the Pleiades cluster, \citet{Micela1996} find typical X-ray luminosities for K-stars of $\log{L_\mathrm{X}}=29.4$~\ergs, which is more than two orders of magnitude above our upper limit for the companion.
We therefore conclude that either the companion has never been an active X-ray source, which seems unusual for a young late-type star, or that the activity of the companion has already dropped to a moderate level. Given the upper limit for the X-ray luminosity, the companion may be an evolved K-type star similar to those found in the solar neighborhood \citep{Schmitt1995}. This also agrees with the upper limit on its rotational velocity of $v\sin{i}<10$~\kms; neglecting the unknown inclination, this would be a value typical for K-type stars on the main-sequence \citep[][p.~389, Table~15.8]{AllenEd4}.

We speculate that the \co \ system, if bound, should be old enough to let the K-star
become sufficiently inactive, while the G-star \coa \ remained more active, maybe through an interaction with its close-in planet. This hypothesis is backed by the recent results
presented by \citet{Brown2011}, who reported on a discrepancy between different age estimations of the host stars of the planetary systems WASP-18 and WASP-19. Both stars harbor a close-in hot Jupiter and appear to be older than attested by their gyrochronological age. \citet{Brown2011} suggested that an inward migration of the hot Jupiters has caused a spin-up of their host stars via tidal interaction. Alternatively, or even additionally, interactions between the planetary and stellar magnetic fields may have reduced the stellar angular-momentum loss as proposed by \citet{Lanza2010}. The \co \ system is among the planetary systems with the shortest orbital periods and should, therefore, be susceptible to these effects.

\subsection{On the dynamics including the companion}

A gravitationally bound stellar companion influences the dynamics of the \co \ system. In particular, it slightly disturbs the planetary orbit. Indeed, \citet{Gillon2010} find a temporal offset of the secondary eclipse, which can be attributed to a slight orbit eccentricity or another interacting body.
Because transit timing variations larger then 10~s are excluded by the CoRoT light curve \citep{Alonso2009}, \citeauthor{Gillon2010} conclude that the planetary orbit has an eccentricity of $\sim0.014$.
Given the present eccentricity of the orbit, the anomalous radius of the planet can be explained by evolutionary scenarios if the models include a third planetary body in the system. \citet{GuillotHavel2011} propose two possible scenarios, which would result in a relatively recent ($\sim20$~Ma) start of the circularization process of the orbit. One requires a planetary encounter and the other is based on the Kozai interaction with a distant body. Our findings clearly favor the latter scenario.

\section{Conclusion}

The \co \ system may be a key to a more profound understanding of the early evolution of planetary systems.
We studied new optical and X-ray data.
Our analysis showed that magnetic activity can be traced through all layers of the stellar atmosphere from the
photosphere to the corona and provided new evidence, helping to answer questions about the age, distance, and evolution
of the system. A detailed analysis of several age indicators showed that an age between $100$ and $300$~Ma is most likely.
Furthermore, we were able to provide an estimate of $270$~pc for the distance of \coa, but with a large
uncertainty.

Beyond answering questions, our analysis also raised new problems. Most notably, the true nature of
\tmass, the optical and potentially physical stellar companion of \coa, remains doubtful.
The apparent presence of a gravitationally bound and, therefore, most likely coeval K-type stellar companion, which,
nonetheless, shows no detectable activity, would challenge our picture of the \co \ system. Either the companion is old enough to have already become inactive, or \coa \ appears to be younger than it actually is. A third body would have a substantial impact on the evolutionary dynamics of the whole system. It may account for the eccentricity of the planetary orbit and may even be responsible for the observed anomalously large radius of \cob.

\begin{acknowledgements}
This work has made use of observational data obtained with UVES at the ESO Very Large Telescope, Paranal, Chile, and the \chan \ X-ray Observatory.
We acknowledge use of the VALD and NIST atomic line databases, DSS, and SIMBAD.
We are indebted to Nikolai Piskunov (University of Uppsala) for making SME available to us.
We are thankful to Eric J. Bubar (University of Rochester) for providing us with his MSPAWN and PYSPEC routines.
We are grateful to Richard O. Gray (Appalachian State University) for making his SPECTRUM line synthesis code publicly available.
We thank Matthias Ammler-von Eiff (Georg-August-Universit\"at G\"ottingen) for sharing his insight with us.
S.S. acknowledges support from the DLR under grant 50OR0703.
S.C. and U.W. acknowledge support from the DLR under grant 50OR0105.
H.M.M. and K.F.H. are supported in the framework of the DFG-funded Research Training Group ''Extrasolar Planets and their Host Stars'' (DFG 1351/1).
\end{acknowledgements}

  \bibliographystyle{aa}
  \bibliography{16961}

\end{document}